\documentclass[onecolumn,12pt]{IEEEtran}
\IEEEoverridecommandlockouts

\usepackage{cite}
\usepackage{amsmath,amssymb,amsfonts}
\usepackage{algorithmic}
\usepackage{graphicx}
\usepackage{textcomp}
\usepackage{xcolor}
\usepackage{multirow}
\usepackage{tabularx}
\usepackage{booktabs}
\usepackage{caption}
\usepackage{multicol}
\usepackage[normalem]{ulem}

\usepackage{times}

\usepackage[pdfstartview=XYZ,
bookmarks=true,
colorlinks=true,
linkcolor=blue,
urlcolor=blue,
citecolor=blue,
pdftex,
bookmarks=true,
linktocpage=true, 
hyperindex=true
]{hyperref}

\usepackage{orcidlink}

\def\BibTeX{{\rm B\kern-.05em{\sc i\kern-.025em b}\kern-.08em
    T\kern-.1667em\lower.7ex\hbox{E}\kern-.125emX}}

\begin{document}

\title{New Machine Learning Approaches for Intrusion Detection in ADS-B\\
}

\author{
\IEEEauthorblockN{ Mikaëla Ngamboé , Jean-Simon Marrocco, Jean-Yves Ouattara, \\José M. Fernandez (Retired), Gabriela Nicolescu } 

\IEEEauthorblockA{
\textit{Computer and Software Engineering}\\
\textit{Polytechnique Montréal}\\
Montréal, Canada \\
}
}

\maketitle

\begin{abstract}

With the growing reliance on the vulnerable Automatic Dependent Surveillance–Broadcast (ADS-B) protocol in air traffic management (ATM), ensuring security is critical. This study investigates emerging machine learning models and training strategies to improve AI-based intrusion detection systems (IDS) for ADS-B. Focusing on ground-based ATM systems, we evaluate two deep learning IDS implementations: one using a transformer encoder and the other an extended Long Short-Term Memory (xLSTM) network, marking the first xLSTM-based IDS for ADS-B. A transfer learning strategy was employed, involving pre-training on benign ADS-B messages and fine-tuning with labeled data containing instances of tampered messages. Results show this approach outperforms existing methods, particularly in identifying subtle attacks that progressively undermine situational awareness. The xLSTM-based IDS achieves an F1-score of 98.9\%, surpassing the transformer-based model at 94.3\%. Tests on unseen attacks validated the generalization ability of the xLSTM model. Inference latency analysis shows that the 7.26-second delay introduced by the xLSTM-based IDS fits within the Secondary Surveillance Radar (SSR) refresh interval (5–12 s), although it may be restrictive for time-critical operations. While the transformer-based IDS achieves a 2.1-second latency, it does so at the cost of lower detection performance.
\end{abstract}


\section{Introduction}
Automated Dependent Surveillance-Broadcast (ADS-B) technology is essential for air traffic management and broadcasting real-time aircraft navigation data\cite{b1}. Its adoption has significantly enhanced flight safety and improved airspace efficiency by enabling better situational awareness for pilots and air traffic controllers. However, ADS-B is vulnerable to cyberattacks\cite{b2,b3,b4}. This vulnerability stems from the absence of entity authentication, data authentication, and data-integrity verification mechanisms in its design.

To address these vulnerabilities, researchers have proposed a series of countermeasures that fall into two main categories: (1) adding an authentication layer to the ADS-B protocol, primarily through cryptographic methods, and (2) detecting altered messages or signals using non-cryptographic techniques such as multilateration, Kalman filtering, physical layer analysis, and machine learning.

One might wonder: If cryptographic methods can effectively prevent intrusions, why is there still a need for non-cryptographic detection techniques? Although cryptographic methods are essential for ensuring message authenticity and are generally effective, they are not infallible. For instance, the theft or misuse of a secret key can compromise the entire protection scheme. In such cases, preventive mechanisms may fail silently, allowing malicious messages to be accepted as legitimate. This is where detection controls become indispensable. Operating on the receiving side, they monitor system behavior to detect signs of data tampering and uncover attacks that bypass or exploit weaknesses in preventive countermeasures. By doing so, detection techniques address residual risks that persist despite strong authentication, thereby complementing cryptographic solutions. Accordingly, we advocate a defense-in-depth strategy that combines multiple layers of security to enhance the resilience of ADS-B against cyber attacks.

In this study, we focus on machine learning strategies for intrusion detection, particularly deep learning methods, due to their effectiveness in addressing anomalies that affect ADS-B data. According to Chandola \emph{et al.} \cite{b5}, an anomaly can be classified as a \emph{point anomaly}, which is a single data point deviating from the expected behavior; a \emph{collective anomaly}, involving data points that together show abnormal behavior; and a \emph{contextual anomaly}, which is anomalous only within specific temporal or operational contexts \cite{b5,b6}. In time-series data, such as ADS-B, anomalies are typically contextual. For example, an aircraft at 10,000 feet may be normal for domestic flights, but anomalous over the Atlantic, where cruise altitudes exceed 30,000 feet.

Traditional clustering methods (e.g., DBSCAN \cite{b7}) and ensemble methods (e.g., Isolation Forests\cite{b8}) effectively detect point anomalies, but struggle with contextual anomalies. Statistical methods rely on distributional assumptions that often fail in practice, and the selection of appropriate test statistics remains challenging \cite{b5}. Deep learning methods, particularly neural networks trained on normal behavioral patterns, better capture the temporal dependencies and multivariate dynamics of ADS-B messages. Studies have highlighted the potential of autoencoders and other deep learning architectures to build robust intrusion detection systems for ADS-B \cite{b9,b10,b11}.

Autoencoders, when combined with recurrent neural networks (RNN \cite{b12}), particularly Long Short-Term Memory (LSTM \cite{b13}) networks, excel in countering coarse attacks, such as jamming \cite{b14}. However, they struggle with subtle message injections, such as gradual attacks, in which a specific feature of the ADS-B message is subtly altered over time. The inherent limitations of LSTM, including irreversible storage decisions and limitations in memory and computational capacity, make it challenging to enhance LSTM-based autoencoder models in this context. This has spurred interest in context-aware architectures such as contextual autoencoders and transformers. Transformers \cite{b15} with their self-attention mechanism, enhance contextual awareness by capturing long-term dependencies more effectively than LSTM. However, self-attention scales quadratically with sequence length, increasing computational and environmental costs. By contrast, extended LSTM (xLSTM) \cite{b16} introduces efficient memory architectures that maintain long-term dependencies through recurrent operations.
This reduces reliance on global attention \cite{b17} and suits time-series data, such as ADS-B, where tracking the order and timing of events is more relevant than accessing the entire context at once. Consequently, xLSTM emerges as a suitable architecture for intrusion detection in ADS\mbox{-}B.

Beyond deep-learning architecture innovations, recent advances in intrusion detection systems (IDS) design have focused on how learning is structured and transferred across tasks \cite{b18}. One promising direction is using transfer learning, where a model is first pre-trained to capture important characteristics of the normal behavior of the system. This learned knowledge is then applied to a downstream anomaly detection task to help the system distinguish between benign and malicious activities more effectively \cite{b18,b19}. By leveraging these pre-trained representations, IDS models can improve generalization and enhance their ability to detect novel or previously unseen attacks \cite{b20}.

With recent advancements in deep learning architectures and IDS implementation strategies, it is essential to assess the potential of emerging solutions to address the ongoing challenge of implementing an efficient IDS for ADS-B. In this study, we propose and evaluate two deep learning-based IDS specifically designed for ADS\mbox{-}B data. Our focus is on ground-based detection systems, such as those used by air traffic management (ATM), where the application of machine learning is less limited by computational constraints than in regular avionics systems. Regarding our IDS,  the first implementation utilizes the encoder component of a transformer architecture, whereas the second is based on the xLSTM architecture. To the best of our knowledge, this is the first implementation of an xLSTM-based IDS specifically tailored to ADS-B. Furthermore, to the best of our knowledge, this is the first application of a transfer learning approach for implementing an IDS for ADS-B.

\begin{figure*}[ht] 
\includegraphics[width=1\textwidth]{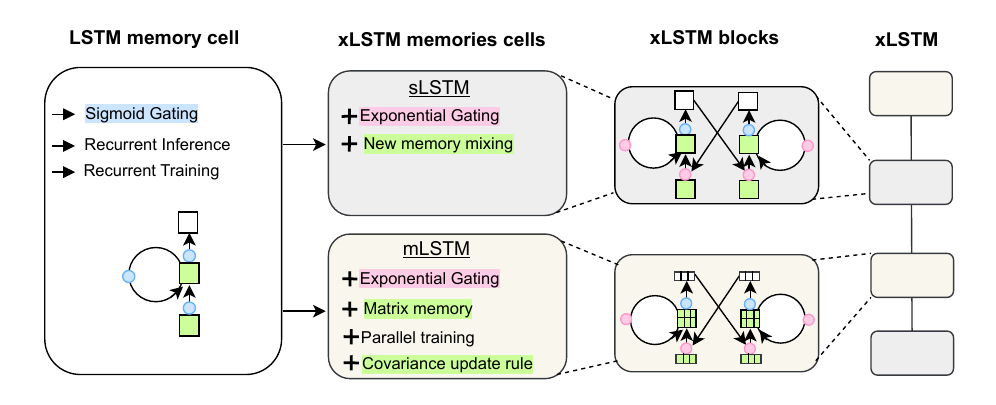}
\caption{Architecture of the original LSTM memory cells and the new xLSTM variants (sLSTM and mLSTM), based on the illustration in paper \cite{b16}}
\label{fig:xlstm}
\end{figure*}

Our transfer learning strategy uses a two-stage process. The models were pre-trained on benign ADS\mbox{-}B traffic to learn contextual patterns and temporal dependencies in ADS-B communications, thereby enhancing generalization across flight trajectories. The pre-trained models were then fine-tuned using labeled datasets to develop specialized models for detecting different types of gradual attacks.  These specialized models were subsequently integrated into a unified multiclass classifier capable of accurately identifying and categorizing various types of ADS-B intrusions. We evaluated the classifier for both known and unknown attacks, and the results show that the xLSTM architecture outperforms the transformer and demonstrates robust generalization to novel threats.

The remainder of this paper is organized as follows. Section~II outlines ADS-B threat models and details the specific one examined in this study. Section~III reviews recent deep-learning-based approaches for intrusion detection in ADS-B. Section~IV provides the background of the extended LSTM (xLSTM) architecture. Section~V details the proposed methodology. Section~VI describes the experimental setup and Section~VII discusses the results. Finally, Section~VIII concludes the paper and outlines directions for future research.

\section{ADS-B Threat Model}

The lack of security controls in ADS-B allows malicious actors to exploit the system by injecting, altering, or suppressing messages without detection \cite{b3,b4}. An overview of the main attacks targeting the ADS-B system is provided in \cite{b2,b3,b4}, including:

\begin{itemize}
    \item \textbf{Eavesdropping:} Passive interception of genuine ADS-B messages.
    \item \textbf{Jamming:} Disruption of RF channels to prevent the transmission of genuine ADS-B messages.
    \item \textbf{Message deletion:} Suppression or removal of genuine ADS-B messages.
    \item \textbf{Message modification:} Alteration of genuine ADS-B messages, potentially falsifying aircraft position, velocity, or identity.
    \item \textbf{Message injection:} Transmission of fabricated ADS-B messages, potentially introducing spoofed aircraft or falsified flight data into the surveillance network.
\end{itemize}

In this study, we deliberately exclude eavesdropping, jamming, and message deletion attacks. Eavesdropping is not considered a direct threat unless combined with active attacks \cite{b3}. Jamming is generic to radio frequency (RF) systems and is typically easy to detect. Similarly, message deletion attacks, when performed independently, are easily flagged due to noticeable gaps in aircraft tracking data.

Our primary focus is on message modification and message injection attacks, which allow adversaries to discreetly alter surveillance data. A significant subset of these attacks is known as gradual attacks, which involve the subtle and continuous alteration of specific ADS-B message features, such as altitude, latitude, etc. These attacks can be executed by either modifying intercepted messages or injecting crafted ones. The insidious nature of gradual attacks enables them to undermine situational awareness over time without triggering immediate alarms, making them particularly difficult to detect and justifying their selection as the central focus of this study.

In this context, the adversary is conceptualized as an unauthorized individual operating from the ground or the air with full control over the communication channel (1090 MHz). This control enables the adversary to suppress legitimate ADS-B traffic, ensuring that the victim (the controller) receives only the information the adversary wishes to convey. This scenario was selected for its plausibility and minimal resource requirements. Tools such as software-defined radios (SDRs) are affordable and widely available, allowing attackers to broadcast falsified ADS-B messages over long distances. Although internal threats from insiders, such as aircraft and airport maintenance technicians, are relevant, this study focuses on external ground-based attacks.

\section{Previous works} 
An overview of the deep learning-driven IDS introduced in the literature to enhance the security of ADS-B is provided below.

Habler and Shabtai \cite{b14} were pioneers in employing machine learning techniques to detect anomalous ADS-B messages through an LSTM encoder-decoder model, which was trained on legitimate flight sequences from takeoff to landing. This model processes new sequences by transforming them into fixed-dimensional vectors using an encoder, followed by reconstruction through the decoder. Anomalous sequences are indicated by higher reconstruction errors. Their approach focuses on data from individual aircraft, overlooking the spatio-temporal correlations among multiple aircraft sharing the same airspace, which compromises accuracy due to limited situational awareness. Akerman \emph{et al.} \cite{b21}  and Olive \emph{et al.} \cite{b22} address this limitation by considering broader traffic flow.

Akerman \emph{et al.}~\cite{b21} aggregates ADS-B messages from multiple aircraft within geographical areas as image streams, using a ConvLSTM encoder-decoder to detect anomalies. The model analyzes image sequences and identifies anomalies when the reconstructed output deviates significantly from the input. An explainability technique provides visual indicators of anomalies to assist in pilot decision-making. Olive  \emph{et al.}~\cite{b22} integrated trajectory clustering with autoencoders to detect anomalies within traffic flows by introducing a custom regularization term based on the distribution distance to optimize the training for sparse clusters. The model generates reconstruction error scores for trajectories, thereby facilitating the identification of anomalous situations in air-traffic operations.

Fried  \emph{et al.} \cite{b23} contend that training distinct models for each location, as demonstrated in \cite{b21,b22}, restricts solutions to flights with sufficient historical data, which is often lacking in business aviation, instructional flying, and aerial work. To address this, they proposed transforming ADS-B data before classification using a non-recurrent autoencoder. These transformations include converting geodetic coordinates to 3D Cartesian coordinates, applying K-lag and K-order differencing to eliminate trends, and extracting time-series characteristics such as maximum, minimum, mean, median, and variance. These properties serve as inputs for a non-recurrent autoencoder. They compared their approach to recurrent autoencoders, noting that their method extracts time-series characteristics, a step omitted by the recurrent autoencoders.

Although the approach in \cite{b23} addresses location-specific model constraints, traditional autoencoders, whether recurrent or non-recurrent, map inputs to fixed points within the latent space, limiting their capacity to capture the full variability of the data. In contrast, models such as variational autoencoders (VAE) present a more flexible alternative by sampling from a distribution defined by the encoder's output, thereby enhancing the representation of uncertainty and variability, which is an advantage for anomaly detection. Luo  \emph{et al.} \cite{b24} proposed a model that integrates a VAE with Support Vector Data Description (SVDD) to detect anomalies in ADS-B data. The VAE is utilized to reconstruct ADS-B messages, and the reconstruction error is employed to train the SVDD model, which establishes a threshold around normal data. During the testing phase, messages that exceeded this threshold were identified as anomalous.

Chevrot  \emph{et al.} \cite{b25} argue that autoencoder architectures employing LSTM and VAE inadequately account for temporal dependencies and assume a Gaussian distribution, resulting in suboptimal performance. Their proposed contextual autoencoder (CAE) employs a single encoder to capture time-dependent patterns and multiple decoders for specific flight phases. The CAE learns normal patterns and calculates anomaly scores for time windows, establishing thresholds based on the 3-sigma rule to distinguish between normal and anomalous data. Luo \emph{et al.} \cite{b26} propose another context-aware architecture using a transformer for sequence reconstruction in their TTSAD model. This model comprises three modules: the temporal convolutional network (TCN) prediction module, which predicts the next value using temporal correlations; the transformer reconstruction module, which reconstructs the sequence to capture long-range dependencies; and the SVDD threshold determination module, which compares reconstructed to real data to detect anomalies.

In general, the F1-scores for detecting subtle, gradual attacks have increased across successive studies. For instance, \cite{b25} documented scores of 0.886 with LSTM-AE, 0.926 with VAE-SVDD, and 0.939 with CAE for the velocity drift attack, whereas TTSAD achieved a score of 0.94 in the same context. These results underscore two significant insights: (1) context-aware architectures currently offer the best performance for detecting sophisticated ADS-B attacks, such as gradual drifts, and (2) despite these gains, the performance of the context-aware approaches proposed in the state-of-the-art remains insufficient for safety-critical systems such as ADS-B.

Among context-aware models, transformers show promise, as evidenced by TTSAD results. However, their quadratic complexity with sequence length limits scalability, prompting interest in alternatives such as extended LSTM (xLSTM), which enhances LSTM while potentially addressing transformer limitations. In this study, we evaluate the effectiveness of the xLSTM architecture in detecting subtle gradual attacks on ADS-B and compare its performance with that of a transformer-based model; specifically, we use the encoder component of the transformer. Our work differs from previous studies in two ways. First, we implement an IDS using the xLSTM architecture, marking its first reported application to ADS-B intrusion detection. Second, we apply transfer learning to train both xLSTM and transformer-based models. Both models are pre-trained to capture normal system behavior, and this knowledge is then transferred to the anomaly detection task to better distinguish benign from malicious activities \cite{b18,b19}. By leveraging these pre-trained representations, the IDS systems can improve generalization and enhance their ability to detect novel or previously unseen attacks\cite{b20}.

\section{Background on xLSTM}
A background on the xLSTM architecture \cite{b16} is provided to support understanding of the proposed approach; readers already familiar with this architecture may skip this section without loss of continuity.

The xLSTM architecture has been designed to elevate the sequence modeling capabilities of LSTM through two main innovations, as depicted in Fig.\ref{fig:xlstm}: exponential gating and new memory structures. Exponential gating enhances the control of information flow within the network by employing more adaptable and stable gating mechanisms, thereby strengthening the ability of the xLSTM model to process and retain relevant data.

Furthermore, xLSTM introduces two advanced memory cell types: scalar LSTM (sLSTM), which employs a refined scalar-based memory update and mixing strategy, and matrix LSTM (mLSTM), which arranges memory cells into matrices, enabling parallel computation and a covariance-based update rule. This matrix-based approach not only expands the memory capacity but also improves the handling of long-range dependencies and complex data patterns. These innovations are embedded within residual block backbones, referred to as xLSTM blocks, and are stacked to form deep xLSTM architectures.

\begin{figure}[h]
    \centering
    \includegraphics[width=0.7\columnwidth]{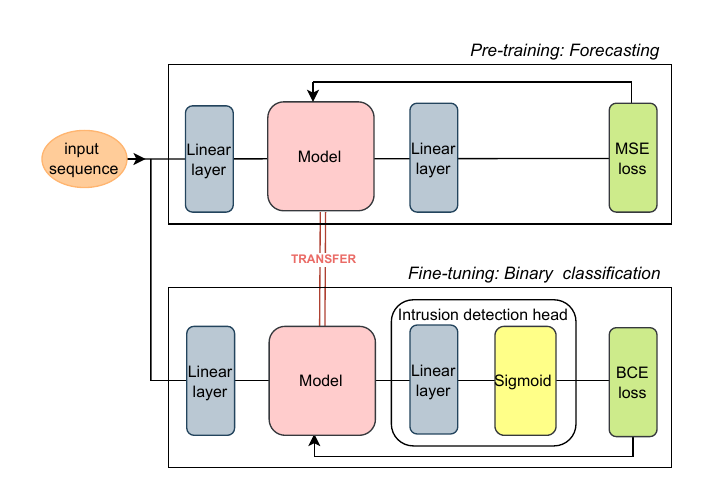}
    \caption{ Methodology for pre-training and fine-tuning. Models are first pre-trained to predict future ADS-B messages by minimizing the mean squared error (MSE) loss. They are then fine-tuned using transfer learning for binary classification tasks, learning to distinguish between benign and malicious traffic by minimizing the binary cross-entropy (BCE) loss.}
    \label{fig:Training_procedure}
\end{figure}

\section{Methodology}
In this section, we outline the procedure for implementing the proposed IDS for ADS-B, followed by a detailed explanation of how the datasets were constructed and attacks were injected into them to train and evaluate the proposed models.

\subsection{IDS Implementation}
To implement the IDS for ADS-B, we adopted a three-step methodology: pre-training, fine-tuning, and multiclass classification.

First, deep learning models were pre-trained in an unsupervised manner to improve their ability to generalize across diverse ADS-B message sequences (\emph{ergo} diverse flight trajectories). This step enabled the model to learn the contextual patterns of ADS-B communications, which is essential for detecting anomalous behavior in dynamic airspace environments. As illustrated in Fig. \ref{fig:Training_procedure}, the input sequence underwent a linear transformation before being passed to the core model. The model was then trained to perform a forecasting task, where it predicted future message values. The prediction passed through another linear layer, and the training objective was to minimize the mean squared error (MSE) loss. As demonstrated in prior studies, unsupervised pre-training positions deep architectures within favorable regions of the parameter space, leading to improved convergence and generalization during supervised learning \cite{b18}.

Second, the pre-trained models were fine-tuned in a supervised manner on specific binary classification tasks, each aimed at detecting a particular class of ADS-B attack. In this phase, we adopted a transfer learning approach, where the model to be fine-tuned was initialized with pre-trained weights and retained the same architecture. As shown in Fig. \ref{fig:Training_procedure}, the input sequence followed a similar preprocessing path to that used during the pre-training phase; however, the output was now fed into an intrusion detection head consisting of a linear layer and a sigmoid activation. The model was then trained to classify traffic as either benign or malicious by minimizing a binary cross-entropy (BCE) loss, which measured the discrepancy between the predicted probabilities and the true labels. This process allowed each fine-tuned model (or binary classifier) to specialize in recognizing the characteristics of a specific attack type.

Finally, we integrated the fine-tuned models into a multiclass classifier capable of simultaneously detecting and categorizing different ADS-B intrusions. Indeed, this final step is crucial for real-world applicability, where network traffic is subject to a variety of intrusion types. By enabling fine-grained threat identification, the multiclass approach supports timely and targeted mitigation strategies, which are vital for maintaining the integrity and safety of air traffic surveillance systems.

\subsection{Data Acquisition and Dataset Implementation} \label{dataset_implementation}
In this work, we used state vector data collected by the OpenSky Network \cite{b27}, a community-based receiver network that continuously gathers air traffic surveillance data for research purposes. State vectors provide an abstraction of tracking information. This data, available in 10-second update intervals, is derived from ADS-B and Mode S messages, offering a summary of the state of an aircraft at a given moment.

We constructed three datasets: Dataset A, Dataset B, and Dataset C, each of which corresponded to a different day and time of data collection to reflect varying flight and navigation conditions. Each dataset was used at a different stage of our methodology: Dataset A for unsupervised pre-training (forecasting), Dataset B for supervised fine-tuning (binary classification), and Dataset C for multiclass classification. The following paragraphs describe the dataset construction process.

Initially, each dataset contained a mix of ADS-B messages from multiple flights. To ensure flight-level coherence, we grouped messages by flight identifier (\emph{callsign}) so that each message sequence corresponded to a single flight. We then discarded flights with missing or incomplete data. Finally, we removed unnecessary fields and retained only the most relevant features for our use case. Namely, aircraft ICAO ID, latitude, longitude, groundspeed, heading, vertical rate, and altitude.

Then, we introduced gradual attacks in Datasets B and C. We focused on this category of attacks because they represent a subtler and more dangerous threat model than abrupt or disruptive attacks (e.g., jamming or replay). While existing models are generally effective at detecting high-noise disruptions, they often fail to detect low-profile message injection attacks that gradually alter flight parameters. Such attacks that may go unnoticed by human operators could have severe consequences.

In a gradual attack, a specific ADS-B message feature is modified incrementally over time: the first message is altered by $\Delta x$, the second by $2\Delta x$, the third by $3\Delta x$, and so forth. In our implementation, we applied three gradual attack types: $+82$ feet per message on altitude, $+1.9$ knots per message on groundspeed, and $+1$ degree per message on heading.

For binary classification, Dataset B was prepared using a one-vs-rest (OvR) strategy. After dividing the dataset into training and test sets, four distinct subsets were derived from each split: \emph{altitude-vs-rest}, \emph{groundspeed-vs-rest}, \emph{heading-vs-rest}, and \emph{benign-vs-rest}. In each subset, 50\% of the flights were subjected to a gradual attack on the target feature and labeled as 1, while the remaining 50\%—including flights affected by other types of attacks or containing only genuine messages—were labeled as 0. This setup enabled each binary classifier to focus on distinguishing a specific attack type from all other conditions, laying the groundwork for the final multiclass classification.

The strategy for constructing Dataset C involved applying each of the previously defined gradual attacks to a portion of the flights while leaving others unaltered. Each class—altitude, groundspeed, heading, and benign—was assigned a unique label. Care was taken to balance the number of samples across all classes to prevent bias during the multiclass training process.

\begin{figure*}[ht] 
\includegraphics[width=1\textwidth]{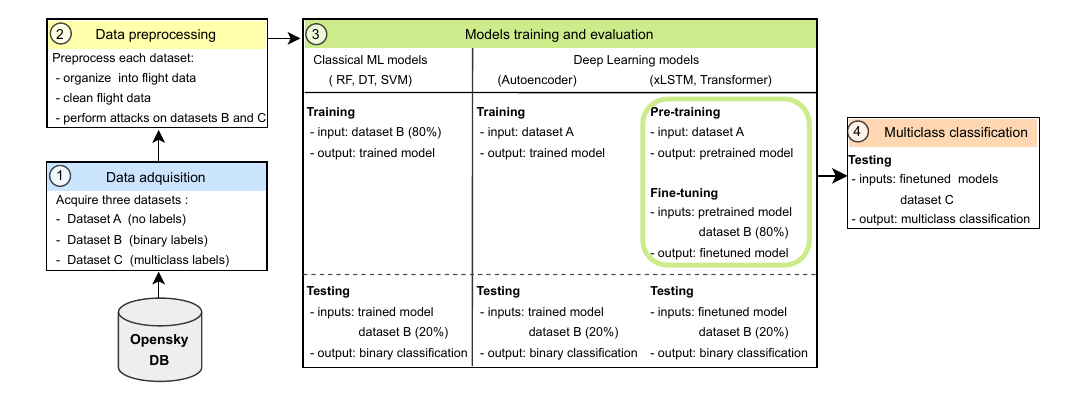}
\caption{Overview of the experimental methodology. Dataset A contains genuine data; Dataset B includes binary-labeled genuine and tampered data; Dataset C has multiclass labels. Classical ML models are trained on Dataset B. The autoencoder is trained on Dataset A and tested on B using reconstruction error. xLSTM and transformer models are pretrained on A and fine-tuned on B. An ensemble of the fine-tuned models performs multiclass classification on Dataset C.}
\label{fig:experiments}
\end{figure*}

\section{Experiments}
Here, we describe the experimental procedure and explain the evaluation methodology used to assess the performance of the proposed IDS. We also describe the hyperparameter optimization process.

\subsection{Experimental Setup} \label{experimentalsetup}
Fig.~\ref{fig:experiments} illustrates the experimental setup used to implement both the xLSTM-based and transformer-based IDS. The process started with the acquisition of three datasets from the OpenSky Network \cite{b27}. Dataset A contained benign, unlabeled ADS-B messages. Dataset B is a hybrid dataset with binary labels indicating either attack or benign traffic. Dataset C is also hybrid, but labeled for multiclass classification, with each attack assigned a distinct label. After the acquisition, the data were preprocessed. This includes structuring the messages into flight sequences and injecting simulated attacks into Datasets B and C, as explained in Section \ref{dataset_implementation}. The training pipeline for the xLSTM and transformer models was organized into three main stages. 

First, the models were pre-trained on Dataset A using a sequence prediction task. The goal is to predict the next ADS-B message based on the sequence of previous messages. This step helps the models capture the temporal dependencies in benign traffic sequences. Next, the pre-trained models were fine-tuned on Dataset B for binary classification. Four separate binary classifiers were trained. Each is specialized for detecting a specific attack or recognizing benign traffic. Dataset B was divided into 80\% for training and 20\% for testing. Finally, a multiclass classifier was built and tested on Dataset C. This dataset includes all attack types, each of which is labeled with a distinct class. When a new sequence is received, it passes through all four binary classifiers. The prediction with the highest probability is selected as the final output of the multiclass classifier.

In parallel, we trained additional models for performance comparisons. These models serve as benchmarks for xLSTM and transformer-based IDS. We selected three classical machine learning algorithms: Random Forest (RF \cite{b28}), decision tree (DT \cite{b29}), and support vector machine (SVM \cite{b30}). Each was trained on 80\% of Dataset B and tested on the remaining 20\% for binary classification. In addition, we trained an autoencoder-based model. It uses Dataset A and follows a forecasting objective similar to the pre-training step of our deep learning models. During the testing, 20\% of Dataset B was used. Sequences were classified based on the reconstruction error.

\begin{table*}[h!]
\centering
\caption{Best hyperparameter configurations for pre-trained models}
\label{tab:hyperparams_pretrainin}
\renewcommand{\arraystretch}{1.1}
\begin{tabular}{l l c c }
\hline
\textbf{Category} & \textbf{Hyperparameter} & \textbf{xLSTM Model} & \textbf{Transfomer Model} \\
\hline
\multirow{2}{*}{Optimizer} 
    & Optimizer & Adam & Adam\\
    & Learning rate & $8.4 \times 10^{-4}$ & $1.3 \times 10^{-4}$ \\
\hline
\multirow{6}{*}{Model} 
    & Embedding size & 64 & 64\\
    & Number of heads & 1 & 1 \\
    & Number of blocks & 4 & --\\
    & Encoder layers & -- & 4\\
    & slstm block at & 1 & --\\
    & Dropout & -- & 0.005\\
\hline
\end{tabular}
\end{table*}

\begin{table*}[h!]
\centering
\caption{Best hyperparameter configurations for fine-tuned models}
\label{tab:hyperparams_finetuning}
\renewcommand{\arraystretch}{1.1}
\begin{tabular}{llcccccccc}
\toprule
\textbf{Category} & \textbf{Hyperparameter} & \multicolumn{4}{c}{\textbf{ xLSTM Models}} & \multicolumn{4}{c}{\textbf{Transformer Models}} \\
\cmidrule(lr){3-6} \cmidrule(lr){7-10}
 & & \textbf{ALT} & \textbf{GS} & \textbf{HDG} & \textbf{GN} & \textbf{ALT} & \textbf{GS} & \textbf{HDG} & \textbf{GN} \\
\midrule
\multirow{3}{*}{General} 
    & Epochs & 5 & 10 & 10 & 15 & 15 & 10 & 10 & 15 \\
    & Batch size & 50 & 40 & 50 & 30 & 50 & 40 & 40 & 30 \\
    & Sequence length & 50 & 50 & 50 & 50 & 50 & 50 & 20* & 50 \\
\midrule
\multirow{1}{*}{Optimizer} 
    & Learning rate & $6 \times 10^{-5}$ & $2 \times 10^{-4}$ & $5 \times 10^{-5}$ & $1 \times 10^{-4}$ & $8.5 \times 10^{-5}$ & $1.5 \times 10^{-5}$ & $4 \times 10^{-4}$ & $1 \times 10^{-4}$ \\
\midrule
\multirow{1}{*}{Model} 
    & Dropout & -- & -- & -- & -- & 0.14 & 0.056 & 0.028 & 0.24 \\
\bottomrule
\end{tabular}
\end{table*}

\subsection{Performance Evaluation Metrics}
We assessed the effectiveness of our IDS across multiple attack scenarios using the confusion matrix, which provides the basis for five key evaluation values:
\begin{itemize}
    \item \textbf{True positive (TP)}: Malicious messages correctly identified as intrusions.
    \item \textbf{False positive (FP)}: Benign messages incorrectly classified as intrusions.
    \item \textbf{True negative (TN)}: Benign messages correctly identified as non-intrusions.
    \item \textbf{False negative (FN)}: Malicious messages that were not detected as intrusions.
\end{itemize}

These values enable the computation of several standard performance metrics, which collectively offer a comprehensive view of the behavior of the models:

\begin{itemize}
    \item \textbf{Precision} measures the fraction of correctly detected intrusions among all intrusion predictions.
\end{itemize}
\begin{align}
\text{Precision} = \frac{TP}{TP + FP}
\end{align}

\begin{itemize}
    \item \textbf{Recall}, also referred to as the true positive rate (TPR), indicates the fraction of actual intrusions that were correctly identified.
\end{itemize}
\begin{align}
\text{Recall} = \frac{TP}{TP + FN}
\end{align}

\begin{itemize}
    \item \textbf{F1-score} is the harmonic mean of precision and recall. A high F1-score reflects the ability of the model to accurately detect intrusions while maintaining a low rate of false alarms.
\end{itemize}
\begin{align}
\text{F1-Score} = 2 \times \frac{\text{Precision} \times \text{Recall}}{\text{Precision} + \text{Recall}}
\end{align}

\begin{itemize}
    \item \textbf{False alarm rate (FAR)}, also known as the false positive rate (FPR), reflects the proportion of benign messages mistakenly flagged as intrusions. This metric is especially critical for intrusion detection in aviation systems, as excessive false alarms can overwhelm operators and compromise decision-making.
\end{itemize}
\begin{align}
\text{FAR} = \frac{FP}{FP + TN}
\end{align}

By considering these metrics together, we obtained a well-rounded evaluation of the ability of our IDS to accurately detect tampered ADS-B messages while minimizing false alarms.

In addition to the detection performance, we also report the inference time of the IDS, defined as the time taken by the IDS to classify a message. It is a key factor in real-world deployments, particularly in time-sensitive environments, such as air traffic surveillance. Although our implementation was not specifically optimized for speed, the same experimental script was used across all the models to ensure fairness. The only variable that changed between the runs was the model being evaluated.

\subsection{Hyperparameter Optimization}
Tables~\ref{tab:hyperparams_pretrainin} and~\ref{tab:hyperparams_finetuning} present the optimal hyperparameter configurations obtained for the pre-training and fine-tuning phases, respectively. These configurations were derived by following the experimental protocol described in Subsection~\ref{experimentalsetup} and by using the Optuna hyperparameter optimization framework \cite{b31}. Optuna leverages Bayesian optimization techniques to explore the hyperparameter space efficiently and identify high-performing combinations. The optimization process was performed on the training sets, with 80\% of the data used for training and 20\% reserved for validation.

In both tables, the hyperparameters are grouped into three main categories: training and evaluation-related parameters (\emph{general}), optimizer-related parameters (\emph{optimizer}), and architecture-specific parameters (\emph{model}). During the pre-training phase, the search focused on identifying the best model and optimizer hyperparameters, whereas general parameters such as batch size (32), sequence length (10), and number of epochs (20) were manually set. As shown in Table~\ref{tab:hyperparams_pretrainin}, the search led to comparable architectural choices for both the xLSTM and transformer models, particularly in terms of the embedding dimension and attention heads.

In the fine-tuning phase, model-specific hyperparameters identified during pre-training were reused, and the search concentrated on optimizing general parameters. As illustrated in Table~\ref{tab:hyperparams_finetuning}, the optimal sequence length was 50 for all models except the transformer  HDG (heading) model, which achieved the best results with a sequence length of 20. However, to ensure consistency during the subsequent multiclass classification task, all the models were fine-tuned using a sequence length of 50.

\section{Results}
In this section, we present the experimental results. We begin by comparing the performance of classical machine learning and deep learning models in distinguishing between genuine and tampered ADS-B messages in a binary classification task. Next, we assess the effectiveness of xLSTM- and transformer-based classifiers, particularly after they have been fine-tuned for specific types of attacks. We then examine how well these models adapt to unknown attacks, meaning attacks that the model was not trained to recognize. Finally, we analyze the inference time of the model and its impact on the situational awareness of controllers. 

\begin{table}[ht]
    \centering
    \footnotesize 
    \caption{Performance results for the binary classification task consisting of distinguishing between genuine and tampered messages.}
    \label{tab:All_models_bin_classification}
    \begin{tabular}{lcccccc}
        \toprule
        \textbf{Metric} & \textbf{SVM} & \textbf{DT} & \textbf{RF} & \textbf{AE} & \textbf{Tx} & \textbf{xLSTM} \\
        \midrule
        Accuracy   & 0.649 & 0.854 & 0.888 & 0.893 & 0.919 & 0.982 \\
        Precision  & 0.613 & 0.856 & 0.881 & 0.890 & 0.913 & 0.980 \\
        Recall     & 0.811 & 0.852 & 0.897 & 0.901 & 0.926 & 0.984 \\
        F1-score   & 0.698 & 0.854 & 0.889 & 0.891 & 0.920 & 0.982 \\
        FPR        & 0.511 & 0.143 & 0.119 & 0.012 & 0.087 & 0.018 \\
        FNR        & 0.189 & 0.147 & 0.102 & 0.099 & 0.074 & 0.016 \\
        \bottomrule
    \end{tabular}
\end{table}

\begin{figure}[htbp]
    \centering
    \includegraphics[width=0.7\linewidth]{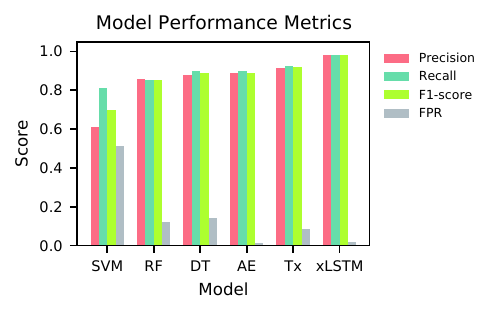}
    \caption{Comparison of performance metrics across six classifiers applied to ADS-B intrusion detection. The xLSTM and transformer models consistently outperform traditional methods, while the SVM exhibits the highest false positive rate.}
    \label{fig:model_metrics}
\end{figure}

\subsection{Binary Classification: Classical vs Deep Learning Models}

Table~\ref{tab:All_models_bin_classification} and Fig.~\ref{fig:model_metrics} show the outcomes of binary classification, where the models are tasked with distinguishing genuine from anomalous ADS-B messages. Deep learning models, particularly the xLSTM and transformer, consistently achieved superior scores across all evaluation metrics. For example, xLSTM achieves a precision of 0.980, a recall of 0.984, and an F1-score of 0.982, whereas the transformer records less impressive but still commendable values of 0.913, 0.926, and 0.920, respectively. Both models, along with the autoencoder, exhibited low false positive rates (FPR). The autoencoder achieved the lowest FPR at 0.012, followed by xLSTM at 0.018, and transformer at 0.087. These findings suggest that deep learning models, particularly xLSTM, are highly robust in identifying sophisticated or stealthy attacks while minimizing false alarms.

The classical machine learning model, Random Forest (RF), also demonstrated respectable performance. With a precision of 0.881 and an F1-score of 0.889, Random Forest appears well suited to scenarios where the nature of attacks is more static or where clear distinctions exist between normal and abnormal patterns. The relatively low FPR of 0.119 further supported this observation. However, not all classical models perform equally well in this regard. For example, the support vector machine (SVM) recorded a significantly higher FPR of 0.511, which may lead to an unmanageable number of false alarms in practice.

These results underscore the strengths of deep learning models in handling complex and nuanced attack scenarios, particularly when the boundary between normal and malicious behavior is subtle. At the same time, they acknowledge the continued relevance of classical machine learning models in more controlled or well-characterized environments. This performance gap ultimately justifies our choice of adopting deep learning architectures for the implementation of the IDS.

\begin{table*}[h]
\centering
\caption{Performance results of the four fine-tuned binary classifiers implemented.}
\label{tab:transformers_lstm_metrics}
\begin{tabular}{lclcccccc}
\toprule
\textbf{Model} & \textbf{Classifier} & \textbf{Target class} & \textbf{Accuracy} & \textbf{Precision} & \textbf{Recall} & \textbf{F1-score} & \textbf{FPR} & \textbf{FNR} \\
\midrule
\multirow{4}{*}{\textbf{xLSTM}} 
 & ALT & altitude     & 0.995 & 0.994 & 0.996 & 0.996 & 0.006 & 0.004 \\
 & GS  & groundspeed  & 0.989 & 0.987 & 0.990 & 0.989 & 0.013 & 0.010 \\
 & HDG & heading      & 0.993 & 0.995 & 0.990 & 0.993 & 0.005 & 0.010 \\
 & BN  & benign       & 0.982 & 0.980 & 0.984 & 0.982 & 0.018 & 0.016 \\
\midrule
\multirow{4}{*}{\textbf{Tx}} 
 & ALT & altitude     & 0.980 & 0.979 & 0.982 & 0.981 & 0.021 & 0.018 \\
 & GS  & groundspeed  & 0.987 & 0.987 & 0.998 & 0.987 & 0.012 & 0.002 \\
 & HDG & heading      & 0.966 & 0.960 & 0.972 & 0.966 & 0.040 & 0.028 \\
 & BN  & benign       & 0.919 & 0.913 & 0.926 & 0.920 & 0.087 & 0.074 \\
\bottomrule
\end{tabular}
\end{table*}

\begin{table}[ht]
\centering
\caption{Performance results of the multiclass classifier when evaluated on unseen data containing known attacks.}
\label{tab:multiclass_over_seendata}
\begin{tabular}{llcc}
\toprule
\textbf{Metric} & \textbf{xLSTM} & \textbf{Transformer} \\
\midrule
 Accuracy   & 0.989 & 0.9432 \\
 Precision  & 0.988 & 0.9434\\
 Recall     & 0.990 & 0.9432\\
 F1-score   & 0.989 & 0.9433\\
 FPR        & 0.012 & 0.056\\
 FNR        & 0.010 & 0.056\\
\midrule
Time (s)    & 7.26& 2.1\\
 \bottomrule
\end{tabular}
\end{table}

\begin{table}[ht]
\centering
\caption{Performance results of the multiclass classifier when evaluated on unseen data containing unknown attacks.}
\label{tab:multiclass_over_unseendata}
\begin{tabular}{llcc}
\toprule
\textbf{Metric} & \textbf{xLSTM} & \textbf{Transformer} \\
\midrule
 Accuracy   & 0.911 & 0.840 \\
 Precision  & 0.920 & 0.853\\
 Recall     & 0.912 & 0.842\\
 F1-score   & 0.910 & 0.832\\
 FPR        & 0.036 & 0.055\\
 FNR        & 0.056 & 0.080 \\
\midrule
Time (s)    & 7.49 & 2.1\\
 \bottomrule
\end{tabular}
\end{table}

\subsection{Fine-Tuning and Multiclass Classification}
Following the binary classification results, we constructed multiclass classifiers using the xLSTM and transformer models. This involved three steps: (1) pre-training the models on genuine ADS-B data, (2) fine-tuning them on labeled samples of specific attacks, and (3) implementing a multiclass classifier using the fine-tuned models. 

Table~\ref{tab:transformers_lstm_metrics} presents the performance of the xLSTM and transformer (Tx) models after fine-tuning. xLSTM consistently outperformed the transformer across all four binary classifiers, achieving high accuracy, precision, recall, and F1-scores, with low false positives and false negatives. Notably, xLSTM achieves an F1-score of 0.982 for benign (BN) samples, indicating a reliable discrimination between normal and malicious behaviors. In contrast, the transformer shows a drop in performance for benign messages, with an F1-score of 0.920 and higher error rates. These results suggest that while both models can capture subtle anomalies, xLSTM is more robust, particularly in identifying benign traffic.

These results highlight the effectiveness of the pretraining and fine-tuning approach for intrusion detection. In \cite{b25}, the authors report F1-scores of 0.886 for LSTM-AE \cite{b14}, 0.926 for VAE-SVDD \cite{b24}, and 0.939 for CAE \cite{b25} when detecting velocity drift attacks. In comparison, the TTSAD \cite{b24} method achieves a slightly higher score of 0.94 under the same conditions. In our study, we refer to the variable called velocity in previous works as ground speed (GS). Focusing on the results of the GS classifier in Table~\ref{tab:transformers_lstm_metrics}, the models based on xLSTM and transformers achieve F1-scores of 0.989 and 0.987, respectively, outperforming previous studies. These findings further confirm the value of combining pretraining with targeted fine-tuning to improve detection performance in ADS-B intrusion detection.

Table~\ref{tab:multiclass_over_seendata} lists the results of full multiclass classification. xLSTM achieves an accuracy of 0.989 and an F1-score of 0.989, maintaining its superior performance. However, the transformer suffers from higher FPR and FNR. This comparatively lower performance with respect to xLSTM suggests that further feature engineering or data preprocessing, such as the encoding embedding technique proposed by the authors of~\cite{b20}, may be beneficial to improve the classification capabilities of the transformer model.

\subsection{Generalization to a Novel Attack}
To evaluate the robustness of our multiclass classifiers or IDS, we introduced a new standing still attack that was not included during training. This attack sets the ground speed (velocity) of the aircraft to zero and freezes its position for a short period of time.
Table~\ref{tab:multiclass_over_unseendata} shows that the xLSTM-based IDS performs adequately, achieving an F1-score of 0.910 and correctly identifying the majority of samples from this previously unseen attack. In contrast, the transformer-based IDS struggles to generalize, with a sharp decline in F1-score.

These findings confirm the capacity of the xLSTM-based IDS to generalize to new threats, making it a reliable candidate for real-time anomaly detection in dynamic airspace environments.

\subsection{Inference Time Analysis and System Performance}

Incorporating a security mechanism into ADS-B, whether cryptographic or non-cryptographic, introduces a safety trade-off: messages are not validated instantly, resulting in an uncertainty delay between their reception and verification. In \cite{b32}, the authors assess this delay by comparing it with the refresh time of Secondary Surveillance Radar (SSR) systems to understand how it could affect the situational awareness of air traffic controllers. Following this approach, we use the SSR refresh time as a reference point to evaluate the operational impact of the inference delays introduced by our IDS.

In ATC, radar systems are essential for tracking aircraft positions and maintaining a safe flight separation. ATC service integrates radar data with other surveillance sources, such as ADS-B, to perform data fusion and build a more accurate and reliable picture of the airspace. Rotating radar systems, including SSR and certain Primary Surveillance Radars (PSR), typically operate at 5–12 revolutions per minute (RPM), yielding refresh intervals between 12 and 5 seconds.

As shown in Table~\ref{tab:multiclass_over_unseendata}, our xLSTM-based multiclass classifier introduces an uncertainty delay of 7.26 seconds. This means that the controllers must wait for more than 7 seconds after receiving an ADS-B message to assess its truthworthiness. In contrast, the transformer-based IDS significantly reduces this delay to approximately 2.1 seconds.

These uncertainty delays have different operational implications, depending on the airspace context. In airports and terminal areas, where controllers often have direct line-of-sight (LOS) to aircraft, longer verification delays may be partially mitigated through visual confirmation. However, controllers at area control centers (ACC), which manage en-route traffic without visual contact, depend entirely on sensor data and are therefore more exposed to the risks introduced by delayed message authentication.

Although the xLSTM-based IDS provides a higher detection rate, its longer uncertainty delay poses limitations in time-sensitive ATC environments. While the 7.26-second delay technically falls within the SSR refresh interval range, it is less suitable where faster decisions are critical. The transformer-based IDS, with its shorter delay, improves timeliness, but does not achieve the same detection performance. As such, the xLSTM model may still be viable in low-density or LOS-supported settings. Recent optimizations of the original xLSTM architecture \cite{b33} have aimed to reduce inference time, and future work should assess whether these updated models can preserve detection performance while improving responsiveness to modern ATC needs.

\section*{conclusion}

This study evaluates emerging solutions for implementing efficient intrusion detection systems (IDS) for ADS-B surveillance technology. We investigate two deep learning-based IDS implementations: a transformer architecture and an extended Long Short-Term Memory (xLSTM) model. Both models are trained using transfer learning to evaluate its effect on their performance and generalization, particularly in detecting subtle and previously unseen attacks. Results show that pretraining and fine-tuning improve detection rates. The xLSTM-based model outperforms the transformer-based model, especially in identifying benign traffic and generalizing to new threats, making it well-suited for real-time anomaly detection. These findings emphasize the importance of low-latency architectures for air traffic control decisions. Although the xLSTM-based IDS achieves higher detection rates, its 7.26-second delay limits its applicability in crowded environments, though it remains suitable for low-density settings with visual confirmation. In contrast, the transformer-based IDS offers shorter inference times but lower detection performance. Future research should explore recent xLSTM optimizations \cite{b33} to improve responsiveness while maintaining accuracy. Additionally, quantum-inspired algorithms may enhance computational efficiency and inference speed.

\vspace{12pt}

\end{document}